\documentclass{elsart}
\usepackage{amsmath}
\usepackage{amsfonts}
\usepackage{amssymb}
\usepackage{graphicx}
\newcommand{\wb}{\omega_{\mathrm{b}}}

\newcommand{\m}{m^*}
\newcommand{\singlefig}{.75\textwidth}

\begin{document}

\begin{frontmatter}

\title{Moving breathers in a bent DNA model}

\author[Informatica]{J Cuevas\thanksref{cor}},
\author[Informatica]{F Palmero},
\author[Informatica]{JFR Archilla},
\author[Fisica]{FR Romero}
\thanks[cor]{Corresponding author.
             E-mail: jcuevas@us.es}

\address[Informatica]{Escuela T\'{e}cnica Superior de Ingenier\'{\i}a
Inform\'{a}tica. Departamento de Fisica Aplicada I. Universidad de
Sevilla. Avda. Reina Mercedes, s/n. 41012-Sevilla (Spain)}

\address[Fisica]{Facultad de Fisica. Universidad de Sevilla.
    Avda. Reina Mercedes, s/n. 41012-Sevilla (Spain)}

\journal{Physics Letters A}

\begin{abstract}

We study the properties of moving breathers in a bent DNA model
with short range interaction, due to the stacking of the base
pairs, and long range interaction, due to the finite dipole moment
of the bonds within each base pair. We show that the movement of a
breather is hindered by the bending of the chain analogously to a
particle in a potential barrier.

\end{abstract}


\end{frontmatter}

\section{Introduction}

Discrete breathers are localised oscillations that appear in
nonlinear discrete systems. Their existence and stability, under
some rather relaxed conditions, was proven by MacKay and Aubry
and, ever since, they have been widely studied
\cite{MA94,A97,FW98}. These localized excitations, under certain
conditions, can move and transport energy along the system and
they are usually called moving breathers
\cite{MER98,MER00,MRE01,CAT96,AC98}.

A particularly interesting discrete system is the Deoxyribonucleic
acid, or DNA. In this system, localization of energy has been
suggested as a precursor of the transcription bubble
\cite{PB89,DPW92,DPB93}, and moving localized oscillations as a
method of transport of information along the double strand
\cite{SK94}.

In order to explain some aspects of DNA dynamics, a great number
of mathematical models have been proposed \cite{Y98}. An
interesting approach has been followed by Peyrard and Bishop (PB
model) \cite{PB89}, who proposed a model in order to study the
dynamics and the thermodinamics of base pairs opening in DNA
denaturation and transcription. In this model, the double strand
is equivalent to a Klein-Gordon chain, the variables are the
distances between nucleotides within each base pair, and only
short range interactions due to the stacking coupling are
considered. Other DNA models ignore this kind of interaction and
only consider long range interactions, whose origin lies in the
dipole moments that characterize the hydrogen bonds between the
nucleotides \cite{MCGJR99,GMCR97}. Nevertheless, we have shown
that the existence of stacking interaction is a necessary
condition to obtain moving breathers in these last kind of models
\cite{CAGR01}.

In the framework of the Peyrard-Bishop model, the bending of the
molecule is not relevant, because only nearest neighbour
interaction is considered. However, when long range interactions
are taken into account, the bending becomes relevant, and it can
modify the dynamics of the system. Bending of DNA has been studied
as an inhomogeneity in a chain with only nearest neighbour
interaction \cite{TP96,FCP97}, and long range interaction has been
taken into account in homogeneous bent DNA models using the
nonlinear Schr\"{o}dinger equation (NLS) \cite{GMC00,CGM00}. The
relationship between static breathers and bending has also been
studied in the framework of the Klein--Gordon equation
\cite{ACMG01,ACG01}.

In this paper, we study the properties of moving breathers in a
bent DNA chain with short and long range interactions in a
modified Peyrard-Bishop model \cite{CAGR01}. The origin of the
interactions lies, respectively, in the stacking of the base pairs
and the finite dipole moment of the hydrogen bonds within each
base pair.

We have found that the bending of the chain can drastically change
the properties of moving breathers. They can cross the bent region
or be reflected, a behaviour which resembles the movement of a
particle in a potential barrier. A similar phenomenon has been
recently observed in a two--dimensional FPU model for rigid
biopolymers \cite{TSI02,IST02}.

Another important result is that the moving breather behaves as a
quasi-particle with constant mass provided that the phonon
radiation is small enough. Under these conditions, we have
developed a method for calculating the instantaneous translational
kinetic energy of the moving breather.

\section{The model} \label{model}

We consider a modification of the Peyrard-Bishop model, which
consists in the addition of an energy term to the Hamiltonian,
that takes into account the long range interaction due to the
dipole--dipole forces \cite{CAGR01}.

The Peyrard--Bishop model has been studied assuming that the
double strand has a planar geometry, i.e., the DNA chain lies in a
plane. In order to study the influence on the dynamics of a
different geometry, we suppose that the plane has been bent and
adapted to a parabola of curvature $\kappa$ (that is, the location
of the n-th base pair is determined by the equation $y_n=\kappa
x_n^2/2$) \cite{ACG01}. All the dipole moments are perpendicular
to the parabola and parallel among them.

Thus, the hamiltonian of the system can be written as:

\begin{equation}\label{ham1}
    H=T+U_{BP}+U_{ST}+U_{DD}
\end{equation}

where $T$ is the kinetic energy:

\begin{equation}
    T=\frac{1}{2}m\sum_n\dot u_n^2,
\end{equation}

being $u_n$ the transverse stretching of the hydrogen bonds
connecting the two bases and $m$ the mass of a nucleotide.

The term $U_{BP}$ represents the interaction energy due to the
hydrogen bonds within each base pair:

\begin{equation}
    U_{BP}=\sum_n V(u_n),
\end{equation}

where $V(u_n)$ is the Morse potential, i.e.,
$V(u)=D(e^{-b\,u}-1)^2$, being $D$ the well depth, which
represents the dissociation energy of a base pair, and $b$ a
spatial scale factor.

$U_{ST}$ is the short range interaction term, representing the
stacking energy between base pairs:

\begin{equation}
\label{stacking}
    U_{ST}=\frac{1}{2}k\sum_n(u_{n+1}-u_n)^2,
\end{equation}

where $k$ is the stacking coupling constant.

$U_{DD}$ is the long range interaction term, due to the
dipole--dipole interaction. It can be expressed as \cite{CAGR01}:

\begin{equation}
\label{dipole}
    U_{DD}=\frac{1}{2}\sum_{n,i}J^*_{ni}u_n u_i,
\end{equation}

where,

\begin{equation}
    J^*_{ni}=\left\{
    \begin{array}{ll}
    \frac{\displaystyle J^*}{\displaystyle |\vec r_n-\vec r_i|^3} &
    \mathrm{for} \; i \neq n \\
    0 & \mathrm{for} \; i=n.\\
    \end{array}
    \right.
\end{equation}

The vector $\vec r_n$ describes the position of the n-th base
pair. We assume that the chain is inextensible, so that the
distance between neighbouring sites remains constant: $|\vec
r_n-\vec r_{n+1}|\equiv d$.

The coupling constant $J$ is related to the charge transfer due to
the formation of the hydrogen bonds ($q$) and the distance between
base pairs ($d$), in the following way:

\begin{equation}
\label{J}
    J^*=\frac{q^2}{4\pi\varepsilon_od^3}.
\end{equation}

The Hamiltonian can be written as:

\begin{equation}\label{ham}
    H=\sum_{n=1}^N\left(\frac{1}{2}m\dot u_n^2+D(e^{-b\,u_n}-1)^2+
    \frac{1}{2}k(u_{n+1}-u_n)^2+
    \frac{1}{2}\sum_{i}J^*_{in}u_iu_n\right),
\end{equation}

With an appropriate change of variables \cite{CAGR01}, the
distance between neighbouring sites is 1 and the dynamical
equations become:

\begin{equation}
\label{F}
   F(\{u_n\})\equiv \ddot u_n+(e^{-u_n}-e^{-2u_n})
   +C(2u_n-u_{n+1}-u_{n-1})+\sum_{i}J_{in}u_i=0,
\end{equation}

where $C=k/2Db^2$ and $J_{in}=J^*_{in}/2Db^2$. Furthermore, we
define $J=J^*/2Db^2$ as a new dimensionless dipole--dipole
coupling constant.

\section{Moving breathers and parameter values}

Our aim is to study the dynamics of a moving breather travelling
through a bent chain. In order to obtain a moving breather, we
perturb the velocity of a static breather far from the parabola
vertex, launching it to the bent part. Static breathers have been
calculated using common methods based on the anticontinuous limit
\cite{MARIN,MA95,MA96,AMM99}. Once a suitable static breather is
obtained, it is made movable using the marginal mode method
\cite{CAT96,AC98}, which basically consists in adding to the
velocities of the static breather a perturbation of magnitude
$\lambda$ colinear to the direction of a linear localized mode,
and letting the system evolve in time. In this context, an useful
concept for describing the moving breather dynamics is the
effective mass \cite{CAT96,AC98}, a measure of the moving breather
inertia. If the kinetic energy added to the breather is
$E=\lambda^2/2$, it is found that the resulting translational
velocity of the breather, $v$, is proportional to $\lambda$
\cite{CAT96}. Consequently, moving breathers can be considered as
a quasi-particle with mass $\m$, which can be defined through the
relation:

\begin{equation}\label{m}
    \frac{1}{2}\m v^2=\frac{1}{2}\lambda^2=E.
\end{equation}

A difficult issue is the choice of appropriate values of the
parameters in order to fulfill two different requirements. On the
one hand, there is only a small range of them that allows breather
mobility. On the other hand, the parameters must be consistent
with real DNA. The last aspect is controversial by itself as
explained below. The requirement that the breathers can move is
our first priority, in agreement with the experimentally observed
``breathing'' modes in DNA \cite{PLZP79,MRZ80}.


While there exists a general agreement in the literature about the
order of magnitude of the parameters $D$ and $b$, there are not
accurate data about the value of the elastic constant $k$, which
oscillates between $0.01 eV/$\AA$^2$ and $10 eV/$\AA$^2$
\cite{DPW92,CM00}. In their original paper, Peyrard and Bishop
\cite{PB89} used a set of values $D=0.33$ eV, $b=1.8$ \AA$^{-1}$
and $k=3.0 \times 10^{-3}$ eV/\AA$^2$. This set of parameters
gives a value of $C=0.014$. Consequently, as $C$ must be greater
than $0.12$ in order to obtain moving breathers \cite{CAGR01},
they would not exist in the Peyrard--Bishop model with these
values of the parameters.

For a given model, a tuning procedure must be achieved in order to
correctly fit the parameters to experimental results. Even the
experimental results are controversial as each one refers to
different aspect of DNA dynamics (torsion, bending, stretching,
...) \cite{BCPR99}. Some recent works propose a generalization of
the PB model with different values of the elastic constant $k$. In
\cite{BCPR99}, the selected parameter values gives a coupling
parameter $C=0.63$. In \cite{CM99,CM00}, normal modes
corresponding to hydrogen bonds excitations are theoretically
investigated, with a range of the coupling parameter
$C\in(0.06,0.31)$ that shows a good agreement with neutron
scattering experiments.

In this paper, we have chosen the breather frequency $\wb=0.8$, so
that the nonlinear effects are significant but not too strong, as
the nonlinearities in DNA are thought to be weak. The stacking and
dipole--dipole coupling parameters chosen are, respectively,
$C=0.24$ and $J=0.02$, which provides with moving breathers with
low phonon radiation for small enough values of $\lambda$
\cite{CAGR01}.  We have considered different spatial configuration
varying the parameter $\kappa$.

\section{Numerical results}

\begin{figure}
  \begin{center}
    \includegraphics[width=\singlefig]{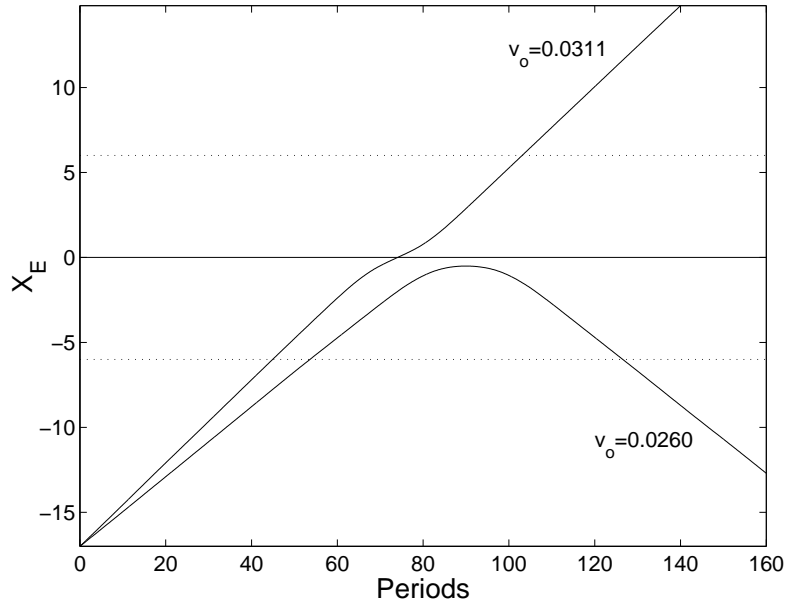}
  \end{center}
  \caption{Evolution of the breather energy centre ($X_E$) for two
  different initial velocities. The curvature is $\kappa=2$.}
  \label{bentcen}
\end{figure}

\begin{figure}
  \begin{center}
    \includegraphics[width=\singlefig]{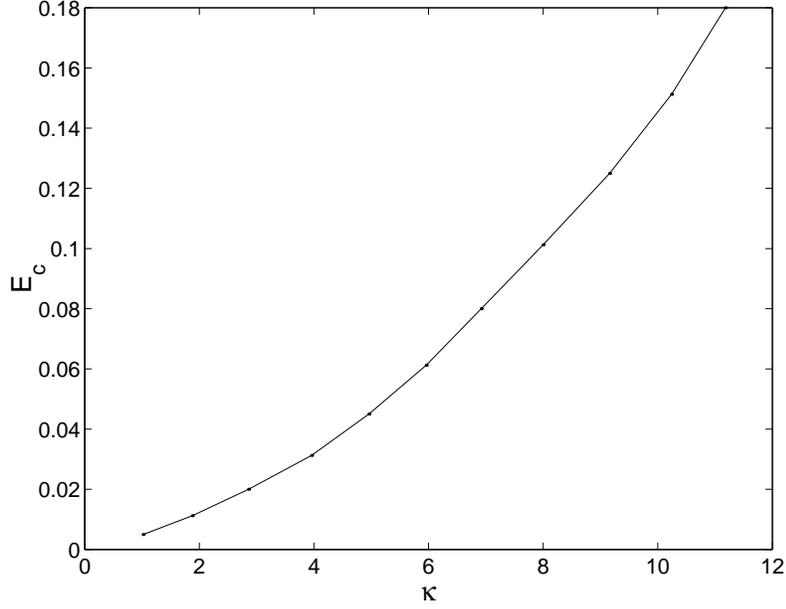}
  \end{center}
  \caption{Critical kinetic energy $E_c$ versus curvature $\kappa$.}
  \label{enkap}
\end{figure}

In figure \ref{bentcen} we show the evolution of the energy centre
\cite{CAGR01} of a moving breather in a bent chain. If the added
kinetic energy, $E=\lambda^2/2$ is smaller than a critical value
$E_c$, the breather rebounds, but, if $E>E_c$, the breather passes
through the bending point. Figure \ref{enkap} shows that the
critical energy increase monotonically with the curvature.

\begin{figure}
  \begin{center}
    \includegraphics[width=\singlefig]{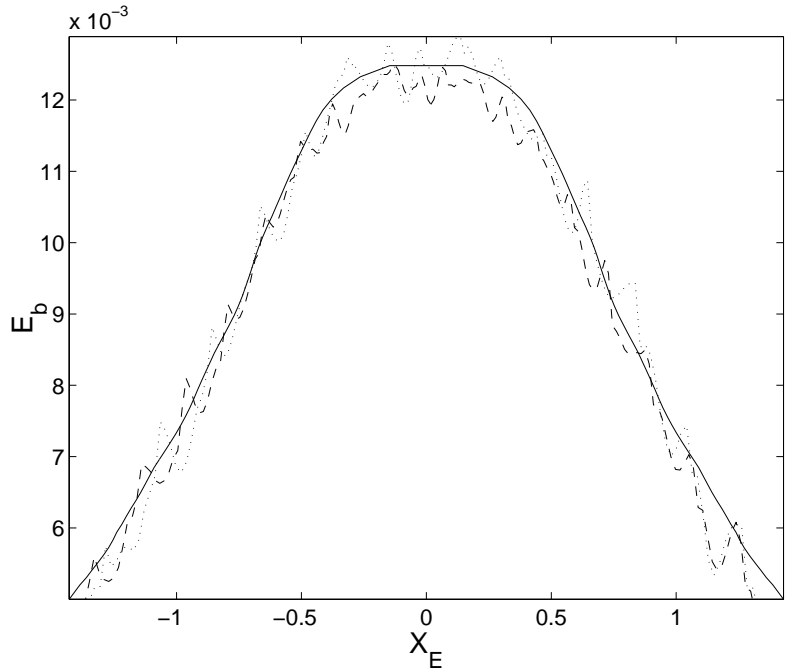}
  \end{center}
  \caption{Potential barrier calculated finding the turning points
  (solid line) and using Eq. (\ref{barrier}) for
  $E=0.0162$ (dashed lined) and $E=0.0200$ (dotted line).
  The curvature is $\kappa=2$ and the critical energy
  is $E_c\approx0.0126$. The zero value of the energy centre
  ($X_E$) represents to the bending point.}
  \label{barrbent}
\end{figure}

Bending acts as a hindrance for the movement of breathers. This
hindrance reminds to the experimented by a particle in a potential
barrier. In this case, breather can be consider as a
quasi-particle and this barrier can be calculated by finding the
points where breathers rebound (or turning points) for different
values of $E$. Furthermore, if the breather has a constant mass
$\m$, this potential barrier can be obtained using the expression:

\begin{equation} \label{barrier}
    E_b=\frac{1}{2}\lambda^2\left(1-\left(\frac{v}{v_o}\right)\right)
\end{equation}

where $v$ is the translational velocity and $v_o$ is its value at
$t=0$. In figure \ref{barrbent} can be observed a good agreement
between the barrier calculated using both methods for a given
value of $\kappa$. The barrier calculated by the second method
exhibits an irregular shape, whose origin lies in the non-uniform
behaviour of the translational velocity due to the discreteness of
the system \cite{Cretegny}. This result confirms that, in this
case, a moving breather behaves as a particle of constant mass
$\m$.

For the parameters used in our study, we have observed that the
effect of the phonon radiation is not negligible if
$E\gtrsim0.03$, and the properties of the moving breathers change.
Thus, in order to have a good agreement between both methods of
calculating the potential barrier, the curvature of the parabola
must be $\kappa\lesssim4$ (see figure \ref{enkap}). Otherwise, the
value of the energy necessary for a moving breather to cross the
bending point will be so high that the phonon radiation will alter
the movement.

\section{Conclusions}

In this paper, we have studied the behaviour of moving breathers
in a bent DNA chain. The movement of a breather is qualitatively
similar to the experienced by a particle in a potential barrier.
For a given initial velocity, breathers can cross the bending
point if the curvature is lower than a critical value. Otherwise,
the breather rebounds. This fact suggest that a slight
conformational changes in the DNA molecule can modify the dynamic
of nonlinear excitations.

Another result is that the breather effective mass remains
constant when the breather cross the bending point as long as the
curvature is small enough. This fact provides with a method to
calculate the instantaneous translational energy, which is given
by $E_{\mathrm{trans}}=\m v^2/2$, being $v$ the translational
velocity of the breather. In a recent work, it is proposed a
method to calculate the translational energy in a $FPU$ chain
based in the asymmetry between the difference between the maxima
of the potential and kinetic energies and the total energy of the
moving breather \cite{IST02}. This analysis can be performed
because the maximum kinetic and potential energies coincide with
the total energy of the static breather, as the interaction
potential is spatially symmetric. However, this method cannot be
applied in our Klein--Gordon chain because the on--site potential
is not spatially symmetric. A further study could consider a
spatially symmetric potential, as the double-well one, in order to
find out whether both methods are equivalent.

\section*{Acknowledgments}
  This work has been supported by the European Commission under
  the RTN project LOCNET, HPRN-CT-1999-00163.

  JC acknowledges an FPDI grant from `La Junta de Andaluc\'{\i}a'.

\newcommand{\noopsort}[1]{} \newcommand{\printfirst}[2]{#1}
  \newcommand{\singleletter}[1]{#1} \newcommand{\switchargs}[2]{#2#1}


\begin{thebibliography}{10}

\bibitem{MA94}
RS~MacKay and S~Aubry.
\newblock Proof of existence of breathers for time-reversible or
  \mbox{Hamiltonian} networks of weakly coupled oscillators.
\newblock {\em \mbox{Nonlinearity}}, 7:1623--1643, 1994.

\bibitem{A97}
S~Aubry.
\newblock Breathers in nonlinear lattices: Existence, linear stability and
  quantization.
\newblock {\em \mbox{Physica D}}, 103:201--250, 1997.

\bibitem{FW98}
S~Flach and CR~Willis.
\newblock Discrete breathers.
\newblock {\em Physics Reports}, 295:181--264, 1998.

\bibitem{MER98}
JL~Mar\'{\i}n, JC~Eilbeck, and FM~Russell.
\newblock Localized moving breathers in a 2-{D} hexagonal lattice.
\newblock {\em Phys. Lett. A}, 248:225--229, 1998.

\bibitem{MER00}
JL~Marin, JC~Eilbeck, and FM~Russel.
\newblock 2-d breathers and applications.
\newblock In PL~Christiansen and MP~Soerensen, editors, {\em Nonlinear Science
  at the dawn of the 21st century}, pages 293--306. Springer, 2000.

\bibitem{MRE01}
JL~Marin, FM~Russell, and JC~Eilbeck.
\newblock Breathers in cuprate-like lattices.
\newblock {\em Phys. Lett. A}, 281:21--25, 2001.

\bibitem{CAT96}
Ding Chen, S~Aubry, and GP~Tsironis.
\newblock Breather mobility in discrete $\phi^4$ lattices.
\newblock {\em Physical Review Letters}, 77:4776, 1996.

\bibitem{AC98}
S~Aubry and T~Cretegny.
\newblock Mobility and reactivity of discrete breathers.
\newblock {\em Physica D}, 119:34, 1998.

\bibitem{PB89}
M~Peyrard and AR~Bishop.
\newblock Statistical mechanics of a nonlinear model for dna denaturation.
\newblock {\em Physical Review Letters}, 62:2755, 1989.

\bibitem{DPW92}
T~Dauxois, M~Peyrard, and CR~Willis.
\newblock Localized breather-like solution in a discrete {K}lein-{G}ordon model
  and application to {DNA}.
\newblock {\em Physica D}, 57:267, 1992.

\bibitem{DPB93}
T~Dauxois, M~Peyrard, and AR~Bishop.
\newblock Dynamics and thermodynamics of a nonlinear model for {DNA}
  denaturation.
\newblock {\em Phys Rev E}, 47:684, 1993.

\bibitem{SK94}
M~Salerno and Yu~Kivshar.
\newblock {DNA} promoters and nonlinear dynamics.
\newblock {\em Phys Lett A}, 193:263, 1994.

\bibitem{Y98}
LV~Yakusevich.
\newblock {\em Nonlinear dynamics of {DNA}}.
\newblock Wiley series in nonlinear sciences. John Wiley \& sons, Chichester,
  1998.

\bibitem{MCGJR99}
SF~Mingaleev, PL~Christiansen, YuB Gaididei, M~Johansson, and
K\O{} Rasmussen.
\newblock Models for energy and charge transport and storage in biomolecules.
\newblock {\em Journal of Biological Physics}, 25:41--63, 1999.

\bibitem{GMCR97}
YuB Gaididei, SF~Mingaleev, PL~Christiansen, and K\O{} Rasmussen.
\newblock Effects of nonlocal dispersive interactions on self-trapping
  excitations.
\newblock {\em Physical Review E}, 55(3):6141--6150, 1997.

\bibitem{CAGR01}
J~Cuevas, JFR Archilla, YuB Gaididei, and FR~Romero.
\newblock Moving breathers in a {DNA} model with competing short and long range
  dispersive interactions.
\newblock {\em Physica D}.
\newblock In press.

\bibitem{TP96}
JL~Ting and M~Peyrard.
\newblock Effective breather trapping mechanism for dna transcription.
\newblock {\em Phys Rev E}, 53:1011, 1996.

\bibitem{FCP97}
K~Forinash, T~Cretegny, and M~Peyrard.
\newblock Local modes and localization in a multicomponent nonlinear lattices.
\newblock {\em Phys Rev E}, 55:4740, 1997.

\bibitem{GMC00}
YuB Gaididei, SF~Mingaleev, and PL~Christiansen.
\newblock Curvature--induced symmetry breaking in nonlinear schr\"{o}dinger
  models.
\newblock {\em Phys Rev E}, 62:R53, 2000.

\bibitem{CGM00}
PL~Christiansen, YuB Gaididei, and SF~Mingaleev.
\newblock Effects of finite curvature on soliton dynamics in a chain of
  nonlinear oscillators.
\newblock {\em J Phys Condens Matter}, 13:1181, 2000.

\bibitem{ACMG01}
JFR Archilla, PL~Christiansen, SF~Mingaleev, and YuB~Gaididei YuB.
\newblock Numerical study of breathers in a bent chain of oscillators with long
  range interaction.
\newblock {\em J Phys A: Math. Gen.}, 34:6363, 2001.

\bibitem{ACG01}
JFR Archilla, PL~Christiansen, and YuB Gaididei.
\newblock Interplay of nonlinearity and geometry in a {DNA}--related,
  klein--gordon model with long--range dipole--dipole interaction.
\newblock {\em Phys Rev E}.
\newblock In press.

\bibitem{TSI02}
GP~Tsironis, JM~Sancho, and M~Iba{\~n}es.
\newblock Localized energy transport in biopolymer models with rigidity.
\newblock {\em Preprint}.

\bibitem{IST02}
M~Iba{\~n}es, JM~Sancho, and GP~Tsironis.
\newblock Dynamical properties of discrete breathers in curved chains with
  first and second neighbors interaction.
\newblock {\em Phys Rev E}.
\newblock In press.

\bibitem{FPM94}
K~Forinash, Peyrard M, and BA~Malomed.
\newblock Interaction of discrete breathers with impurity modes.
\newblock {\em Phys Rev E}, 49:3400, 1994.

\bibitem{MARIN}
JL~Marin.
\newblock {\em {I}ntrisic {L}ocalized {M}odes in nonlinear lattices}.
\newblock Ph{D} thesis, Universidad de {Z}aragoza, Departamento de {F}ísica de
  la {M}ateria {C}ondensada, June 1997.

\bibitem{MA95}
JL~Marin and S~Aubry.
\newblock Breathers in nonlinear lattices: Numerical methods based on the
  anti--integrability concept.
\newblock In L~V\'{a}zquez, L~Streit, and VM~P\'{e}rez-Garc\'{\i}a, editors,
  {\em Nonlinear Klein--Gordon and Schr\"{o}dinger Systems: Theory and
  Applications}, pages 317--323. World Scientific, Singapore and Philadelphia,
  1995.

\bibitem{MA96}
JL~Marin and S~Aubry.
\newblock Breathers in nonlinear lattices: Numerical calculation from the
  anticontinuous limit.
\newblock {\em \mbox{Nonlinearity}}, 9:1501--1528, 1996.

\bibitem{AMM99}
JFR Archilla, RS~MacKay, and JL~Marin.
\newblock Discrete breathers and anderson modes: two faces of the same
  phenomenon?
\newblock {\em \mbox{Physica D}}, 134:406--418, 1999.




\bibitem{PLZP79}
EW Prohofsky, KC Lu, LL van Zandt, and BF Putnam.
\newblock Brathing modes and induced resonant melting of the
double helix
\newblock{\em \mbox{Phys. Lett. A}}, 70:492-494, 1979.

\bibitem{MRZ80}
DP Millar, RJ Robbins, and AH Zewail.
\newblock Direct observation of the torsional dynamics of {DNA} and
{RNA} by picosecond spectroscopy
\newblock{\em \mbox{Proc. Natl. Acad. Sci. USA}}, 77:5593-5597, 1980.

\bibitem{CM00}
S Cocco and R Monasson.
\newblock Theoretical study of collective modes in DNA at ambient
temperature
\newblock{\em \mbox{J. Chem. Phys.}}, 112:10017-10033, 2000.

\bibitem{BCPR99}
M Barbi, S Cocco, M Peyrard, and S. Ruffo.
\newblock A Twist opening model for DNA
\newblock{\em \mbox{Journal of Biological Physics}}, 24:358-369, 1999.

\bibitem{CM99}
S Cocco and R Monasson.
\newblock Statistical Mechanics of torque induced denaturation of
DNA
\newblock{\em \mbox{Phys. Rev. Lett.}}, 83:5178-5180, 1999.

\bibitem{Cretegny}
T~Cretegny.
\newblock {\em Dynamique collective et localisation de l'\'energie dans les
  reseaux non-linéaires}.
\newblock PhD thesis, \'Ecole Normale Sup\'erieure de Lyon, 1998.



\end{thebibliography}
\end{document}